\documentclass[12pt,preprint2]{aastex}

\slugcomment{To be submitted to \emph{Astrophysical Journal}}

\begin{document}

\title{The Synchrotron Emission Mechanism in the Recently Detected Very High
Energy Radiation from the Crab Pulsar}

\author{Machabeli George\altaffilmark{1} and Osmanov Zaza\altaffilmark{2}}
\affil{Georgian National Astrophysical Observatory, Chavchavadze
State University, Kazbegi 2a, 0106, Tbilisi, Georgia}
\altaffiltext{1}{g.machabeli@astro-ge.org}

\altaffiltext{2}{z.osmanov@astro-ge.org}

\begin{abstract}

Interpretation of the recently discovered very high energy (VHE)
pulsed emission from the Crab pulsar is presented. By taking into
account the fact that Crab pulsar's radiation for the optical and
VHE spectrum peak at the same phases, we argue that the source of
this broad band emission is spatially localized. It is shown that
the only mechanism providing the results of the MAGIC Cherenkov
telescope, should be the synchrotron radiation. We find that in the
magnetospheric electron-positron plasma, due to the cyclotron
instability, the pitch angle becomes non-vanishing, which leads to
the efficient synchrotron mechanism, intensifying on the light
cylinder lengthscales. We also estimate the VHE radiation spectral
index to be equal to $-1/2$.
\end{abstract}

\keywords{instabilities - plasmas - pulsars: individual (PSR
B0531+21) - radiation mechanisms: non-thermal}
\section{Introduction}

One of the fundamental problems concerning pulsars relates the
origin of the high energy electromagnetic radiation. According to
the standard approach, two major mechanisms govern the high energy
radiation: the inverse Compton scattering \citep{blan} and the
synchrotron emission \citep{pacini,shkl}.  On the other hand, up to
now in most of the cases it is not clear where the location of the
high-energy electromagnetic radiation is: closer to the pulsar
(polar cap model, see for example \cite{stur}) or farther out in the
magnetosphere (outer gap model, see for example
\cite{cheng1,cheng2}). An exception is the high energy emission
recently detected by the MAGIC Cherenkov telescope \citep{magic},
which has revealed that the pulsed radiation above $25GeV$ is
inconsistent with the polar cap models. In the framework of these
models, over the star's surface there is a vacuum gap with the
electric field inside \citep{rudsuth}, which accelerates particles
up to relativistic energies leading to the emission process.
Unfortunately, energies of particles accumulated in the gap, are not
enough to explain the observed radiation. To solve this problem
several mechanisms have been proposed. For increasing the gap size,
by \cite{usov} the formation of positronium (electron-positron bound
state) was considered. Another mechanism, leading to the enlargement
of the gap zone was introduced by \cite{arons} and the approach was
based on a process of rectifying of the magnetic field lines. This
method was applied by \cite{hard} for studying the high altitude
radiation from the pulsar slot gaps. The authors consider a
three-dimensional model of optical and $\gamma$-ray emission from
the slot gap accelerator of a spin-powered pulsar and predict that
the slot gap emission below $200MeV$ will exhibit correlations in
time and phase with the radio emission. A general relativistic
approach has been proposed by \cite{musl}, where, taking into
account the fact that in the vicinity of the neutron star, the
space-time is curved, the authors applied the Kerr metric. It was
shown that the gap size increases due to the general relativistic
effects. All aforementioned mechanisms provide the required increase
in the gap area, but it is not enough for explaining the observed
radiation. These problems provoked a series of works considering the
so-called outer gap models, where for studying emission from
pulsars, several mechanisms have been proposed: the inverse Compton
scattering, curvature radiation and the synchrotron emission. On the
other hand, according to the MAGIC Cherenkov telescope, the
observational evidence confirms that the Crab pulsar's pulsed
emission in the optical and in the high energy radiation is
generated at the same location.

In this context the recent detection of VHE gamma rays from the Crab
pulsar could be very important \citep{magic}. From 2007 October to
2008 February, the MAGIC Cherenkov telescope has discovered pulsed
emission above $25 GeV$. Comparing signals in different energy band,
it is shown that for $E>25 GeV$ the signal peaks at the same phase
as the signal measured by MAGIC \citep{magic} for the optical band.
On one hand this indicates that the polar cap models must be
excluded from possible mechanisms of the detected radiation. On the
other hand analysis of the MAGIC data implies that the location of
the high energy emission is the same as that of the low-energy one
\citep{manch}.

In the present paper, referring to the recently published data of
MAGIC, we interpret the results and argue that the observed high
energy radiation is produced by the synchrotron mechanism. For this
purpose we use the approach developed in \citep{machus1} and apply
the method to the Crab pulsar.

The paper is organized as follows. In Section 2, we consider the
synchrotron radiation of electrons, in Sect. 3, we present our
results, and in Sect. 4 we summarize them.

%
\section{Synchrotron emission}
%

According to the theory of synchrotron emission, a relativistic
particle, moving in the magnetic field emits electromagnetic waves
with the following photon energies \citep{Lightman}:
\begin{equation}
\label{eps} \epsilon\approx 1.2\times 10^{-17}B\gamma^2\sin\psi
(GeV),
\end{equation}
where $B$ is the magnetic induction, $\gamma$ is the Lorentz factor
and $\psi$ is the pitch angle. On the other hand, for typical
magnetospheric parameters the timescale of the transit to the ground
Landau state is so small that almost from the very beginning of
motion electrons move quasi-one-dimensionally along the field lines
without emission. The situation changes due to the Cyclotron
instability of the electron-positron magnetospheric plasma, which
"creates" certain pitch angles, leading to the subsequent emission
process \citep{machus1}. In this letter we consider the plasma
composed of two components: (1) the plasma component with the
Lorentz factor, $\gamma_p$ and (2) the beam component with the
Lorentz factor, $\gamma_b$. According to the approach developed by
\cite{machus1}, due to the quasi-linear diffusion the following
transverse and longitudinal-transversal waves are generated
\begin{equation}\label{disp1}
\omega_t =
kc\left(1-\frac{\omega_p^2}{4\omega_B^2\gamma_p^3}\right),
\end{equation}
\begin{equation}\label{disp2}
\omega_{lt} =
k_{_{\|}}c\left(1-\frac{\omega_p^2}{4\omega_B^2\gamma_p^3}-\frac{k_{\perp}^2
c^2}{16\omega_p^2\gamma_p}\right).
\end{equation}
Here, $k$ is the modulus of the wave vector, $k_{_{\|}}$ and
$k_{\perp}$ are the wave vector's longitudinal (parallel to the
background magnetic field) and transverse (perpendicular to the
background magnetic field) components. respectively, $c$ is the
speed of light, $\omega_p \equiv \sqrt{4\pi n_pe^2/m}$ is the plasma
frequency, $\omega_B\equiv eB/mc$ is the cyclotron frequency, $e$
and $m$ are electron's charge and the rest mass, respectively, and
$n_p$ is the plasma density.

For the generation of the aforementioned modes, the cyclotron
resonance condition \citep{kmm}:
\begin{equation}\label{cycl}
\omega - k_{_{\|}}V_{_{\|}}-k_xu_x\pm\frac{\omega_B}{\gamma_b} = 0,
\end{equation}
has to be satisfied. Here, $u_x\equiv
cV_{_{_{\|}}}\gamma_b/\rho\omega_B$ is the drift velocity of
resonant particles, $V_{_{\|}}$ is the component of velocity along
the magnetic field lines and $\rho$ is the curvature radius of field
lines.

When a particle moves along a curved magnetic field line it
experiences a force that is responsible for the conservation of the
adiabatic invariant, $I = 3cp_{\perp}^2/2eB$ \citep{landau}. The
transverse and longitudinal components of the aforementioned force
are given by the following expressions:
\begin{equation}\label{g}
G_{\perp} = -\frac{mc^2}{\rho}\gamma_b\psi,\;\;\;\;\;G_{_{\|}} =
\frac{mc^2}{\rho}\gamma_b\psi^2.
\end{equation}
Since the particle emits ($\lambda < n_p^{-1/3}$) in the synchrotron
regime, the corresponding radiative force will appear
\citep{landau}:
$$F_{\perp} = -\alpha\psi(1 + \gamma_b^2\psi^2),$$
\begin{equation}\label{f}
F_{_{\|}} = -\alpha\gamma_b^2\psi^2,\;\;\;\;\;\alpha =
\frac{2}{3}\frac{e^2\omega_B^2}{c^2}.
\end{equation}
On the other hand, if one assumes Eqs. (\ref{disp1}-\ref{cycl}),
then, one can show that for all frequencies in the optical range
($\sim 10^{15}Hz$), the development of the cyclotron instability
occurs from the distances of the order of the light cylinder radius,
$R_{lc}$ \citep{machus1}. We have used the parameters: $P\approx
0.033 s$ $R_s\approx 10^6cm$, $n_{ps}\approx 1.4\times
10^{19}cm^{-3}$, $B_s\approx 7\times 10^{12}G$ and $\gamma_b\approx
10^8$. Here, $P$ is the pulsar's period, $R_s$ - its radius, and
$n_{ps}$ and $B_s$ are the plasma density and the magnetic field
induction, respectively, close to the star.

These two forces (${F}_{\perp}$ and ${G}_{\perp}$) tend to decrease
the pitch angle of the particle. Contrary to this, the quasi-linear
diffusion, arising through the influence of the generated waves back
on the particles, tries to widen a range of the pitch angles. The
dynamical process saturates when the effects of the above mentioned
forces are balanced by the diffusion. For $\gamma\psi\gg 1$ it is
easy to show that for typical magnetospheric parameters the forces
satisfy $G_{\perp}\ll F_{\perp}$ and $G_{_\parallel}\ll
F_{_\parallel}$. Then, assuming the quasi-stationary case
($\partial/\partial t = 0$), the corresponding kinetic equation can
be presented in the following way \citep{malmach}:
$$\frac{\partial\left[ F_{_\parallel}f\right]}{\partial
p_{_\parallel}} +
\frac{1}{p_{_\parallel}\psi}\frac{\partial\left[\psi
F_{\perp}f\right]}{\partial\psi}
=\frac{1}{\psi}\frac{\partial^2}{\partial\psi\partial
p_{_\parallel}}\left(D_{\perp_{\parallel}}\frac{\partial
f}{\partial\psi}\right)+$$

\begin{equation}\label{kinet}
+\frac{1}{\psi}\frac{\partial}{\partial\psi}\left[\psi\left
(D_{\perp\perp}\frac{\partial}{\partial\psi} +
D_{\perp_{\parallel}}\frac{\partial}{\partial
p_{_\parallel}}\right)f\right],
\end{equation}
where $f = f(\psi,p{_\parallel})$ is the distribution function of
particles, $p_{_\parallel}$ is the longitudinal momentum,
\begin{equation}\label{dif}
D_{\perp\perp}\approx -\frac{\pi^2 e^2n_bc}{2\omega},\;\;\;\;\;\;
D_{\perp_{\parallel}}\approx \frac{\pi^2
e^2n_b\omega_B}{2mc\gamma^2\omega^2},
\end{equation}
are the diffusion coefficients and $n_b = \frac{B}{Pce}$ is the
density of the beam component. By expressing the distribution
function as $\chi(\psi)f(p_{_\parallel})$, one can solve Eq.
(\ref{dif}) \citep{malmach1,nino}:

\begin{equation}\label{chi} \chi(\psi) = C_1e^{-A\psi^4},
f(p_{_\parallel}) = \frac{C_2}{\left(\alpha\bar{\psi}^2\gamma_b^2
-\frac{\pi^2e^2\bar{\psi}n_bc}{\gamma_b}\right)},
\end{equation}
where
\begin{equation}\label{A}
A\equiv \frac{4e^6B^4P^3\gamma_p^4}{3\pi^3m^5c^7\gamma_b},
\end{equation}
and
\begin{equation}\label{pitch}
\bar{\psi}
 = \frac{\int_{0}^{\infty}\psi\chi(\psi)d\psi}{\int_{0}^{\infty}\chi(\psi)d\psi}
\approx \frac{0.5}{\sqrt[4]{A}}.
\end{equation}
is the mean value of the pitch angle. In the expression of
$f(p_{_\parallel})$ that is responsible for the synchrotron
radiation spectrum (see discussion), the first term of the
denominator comes from the synchrotron reaction force, and the
second term is the contribution of the quasi-linear diffusion.

After combining Eq. (\ref{pitch}) with Eq. (\ref{eps}) one gets the
following expression of energy of the synchrotron photons:
\begin{equation}
\label{eps1} \epsilon(GeV)\approx 6\times
10^{-18}\left(\frac{3\pi^3m^5c^7\gamma_b^3}{4P^3e^6\gamma_p^4
}\right)^{\frac{1}{4}}.
\end{equation}
%

\section{Discussion} \label{sec:discus}

As we have discussed, a typical distance where the instability
developes, is of the order of the light cylinder radius, $\sim
10^8cm$. In this area, due to the quasi-linear diffusion, the pitch
angle (see Equation (\ref{pitch})) is created, leading to the
synchrotron radiation.

\begin{figure}
  \resizebox{\hsize}{!}{\includegraphics[angle=0]{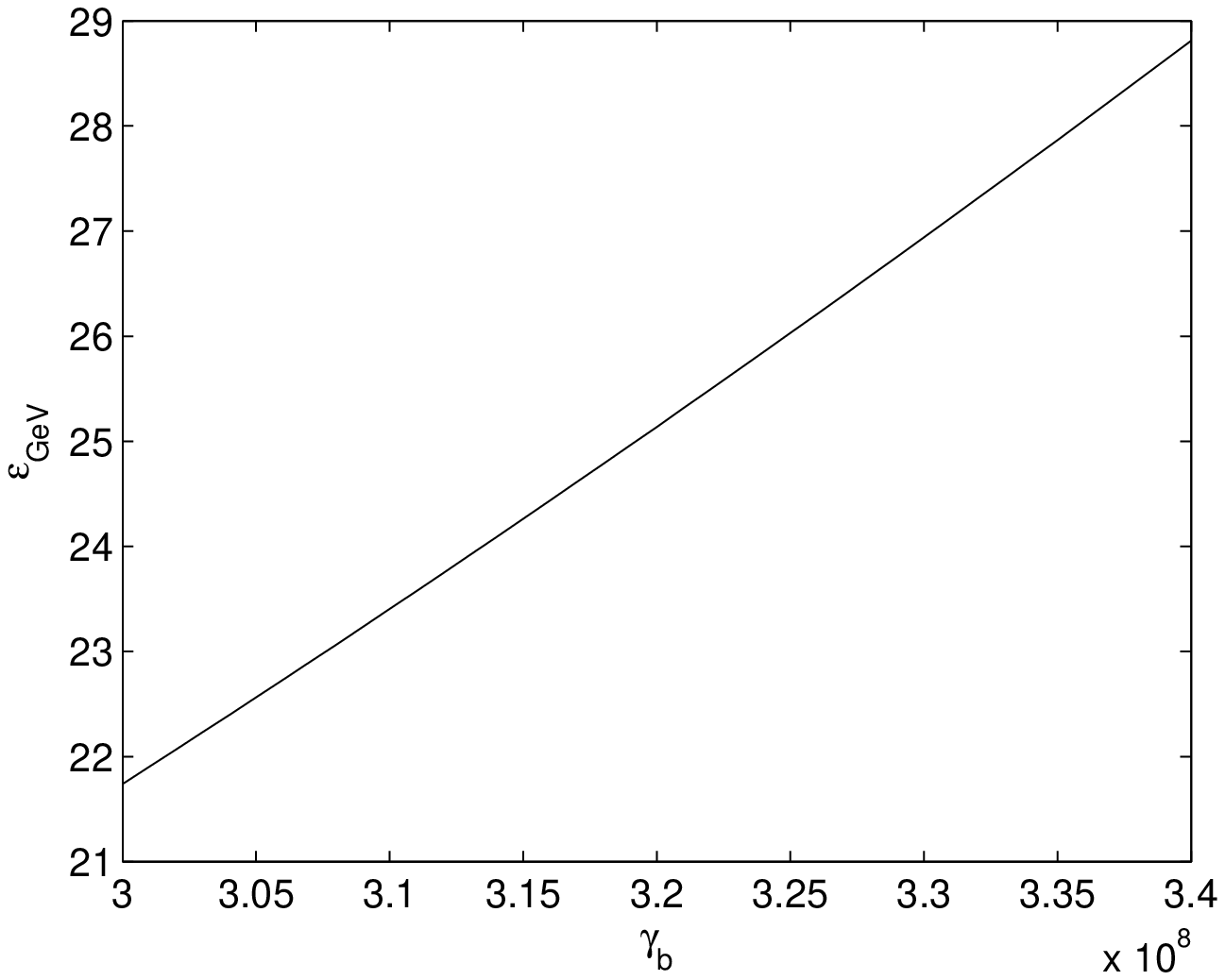}}
  \caption{Behavior of the emission energy vs. the Lorentz
  factor of beam component electrons.
  The set of parameters is $R_s\approx 10^6cm$, $B_s\approx 7\times 10^{12}G$ and
  $\gamma_p \approx 3$.}\label{energy}
\end{figure}

Let us consider Eq. (\ref{eps1}) for the Crab pulsar and assume that
the Lorentz factor of the plasma component is of the order of $\sim
3$ \citep{machus2}. In Figure \ref{energy}, we show the dependence
of the emission energy on the Lorentz factor of the beam component.
The set of parameters is $R_s\approx 10^6cm$, $B_s\approx 7\times
10^{12}G$ and $\gamma_p \approx 3$. As is seen from the figure, the
high energy emission of the order of $25GeV$ is possible for
$\gamma_b\approx 3.2\times 10^8$. This in turn implies that the gap
models providing the Lorentz factors $\sim 10^7$, are not enough to
explain the detected pulsed emission of the Crab pulsar. One of the
possibilities could be the centrifugal acceleration of particles,
when due to the frozen-in condition, electrons move along the
co-rotating magnetic field lines and accelerate centrifugally
\citep{mr,r03,osm7}. Another alternative mechanism for explaining
the observed high energy radiation, could be a collapse
\citep{arcim,zax} of the centrifugally excited unstable Langmuir
waves \citep{incr1} in the pulsar's magnetosphere. Such a
possibility was shown by \cite{mlvm,mvm} for the electron-positron
plasma.

If we take $\gamma_b\approx 3.2\times 10^8$ into account, after
substituting all necessary parameters into Eq. (\ref{pitch}), one
can show that the created (via the quasi linear instability) pitch
angle is of the order of $10^{-5}$. This in turn, confirms our
assumption $\gamma_b\psi\gg 1$, which has been used for constructing
and solving Eq. (\ref{dif}). The timescale of the synchrotron
emission is still very small and the electrons pass very soon to the
ground Landau state and therefore, the distribution function becomes
one-dimensional. Such a distribution function is unstable against
the anomalous Doppler effect and excites the optical radiation. This
mode simultaneously leads to the quasi-linear diffusion, which,
creating the pitch angles, produces the high energy ($>25GeV$)
emission. The synchrotron radiation reaction force limits the pitch
angles which saturate due to the balance between the mentioned
forces and the corresponding diffusion effects. On the other hand,
the credibility of the present mechanism comes from the
observationally evident coincidence of optical and high energy
spectrum phases.

We see that the synchrotron emission can explain the detected
coincidence of the optical and VHE emission phases in terms of the
quasi-linear diffusion. Although, it is supposed that apart from the
synchrotron process, the inverse Compton mechanism and the curvature
radiation can also be responsible for the emission in the pulsar
magnetospheres. But, if this is the case, then, the area of the
inverse Compton and the curvature radiation must be stretched and
not localized, leading to relative shifts of phases, contrary to the
observational pattern.

Expression of $f(p_{_\parallel})$ gives a possibility to predict the
observed spectrum of the synchrotron emission. Indeed, one can
easily show that for Crab pulsar's magnetospheric parameters with
$\gamma_b\sim 3.2\times 10^8$ and $\psi\sim 10^{-5}$, one has
$\alpha\bar{\psi}^2\gamma_b^2\gg\pi^2e^2\bar{\psi}n_bc/\gamma_b$.
Then the distribution function behaves as
$f(p_{_\parallel})\propto\gamma_b^{-2}$. On the other hand,
according to the well known power law formula, the spectrum of the
synchrotron radiation is given by
$I_{\nu}\propto\nu^{-\frac{\beta-1}{2}}$ \citep{ginz}, where $\beta$
describes the particle distribution function,
$f\propto\gamma^{-\beta}$. Therefore, in our case ($\beta = 2$) the
spectral index of synchrotron emission is $-1/2$.

\section{Summary}\label{sec:summary}

\begin{enumerate}

      \item Considering the recently detected VHE emission from the
      Crab pulsar, we studied the role of the synchrotron mechanism in
      producing the observed high energies.

      \item We emphasized that due to very small
      cooling timescales, particles rapidly transit to the ground
      Landau state preventing the subsequent radiation.
      The situation changes thanks to the cyclotron instability,
      which at a certain distance from the star's surface
      develops efficiently and creates non-vanishing pitch angles,
      leading to the efficient synchrotron emission process with
      the following spectral index $-1/2$.

      \item Since the cyclotron instability generates the optical
      spectrum and provokes the increase of the pitch angle, we argue that
      the emission in the aforementioned low (optical) and high ($25GeV$)
      energy intervals originates from well localized regions,
      leading to the observational fact that the signals peak with the same phases.
      This in turn means that the inverse Compton scattering and
      the curvature radiation must be excluded from the consideration and the only
      mechanism providing VHE gamma-rays is the synchrotron
      process.

      \end{enumerate}

\section*{Acknowledgments}
The authors are grateful to an anonymous referee for valuable
comments. The research was supported by the Georgian National
Science Foundation grant GNSF/ST06/4-096.

\end{document}